\documentclass[final,5p,times,twocolumn,authoryear]{elsarticle}

\usepackage{amssymb}
\usepackage{graphicx}
\usepackage{txfonts}
\usepackage{array}
\usepackage{amssymb}
\usepackage{natbib}
\usepackage{lscape}
\usepackage{rotating}
\usepackage{longtable}

\journal{New Astronomy}

\begin{document}

\begin{frontmatter}

\title{Evolution of the disk of $\pi$\,Aqr: from near-disappearance to a strong maximum}

\author[1]{Ya\"el Naz\'e\corref{fnrs}}
\ead{ynaze@uliege.be}
\cortext[fnrs]{Corresponding author, FNRS research associate}
\author[1]{Gregor Rauw}
\author[2]{Joan Guarro Fl\'o}
\author[3]{Arnold De Bruin}
\author[4]{Olivier Garde}
\author[5]{Olivier Thizy}
\author[6]{Franck Houpert}
\author[7]{Ernst Pollmann}
\author[8]{Carl J. Sawicki}
\author[9]{Marco Leonardi}
\author[10]{Malin Moll}
\author[11]{Christoph T. Quandt}
\author[12]{Paolo Berardi}
\author[13]{Tim Lester}
\author[14]{Patrick Fosanelli}
\author[15]{Andr\'e Favaro}
\author[16]{Jean-No\"el Terry}
\author[17]{Keith Graham}
\author[18]{Benjamin Mauclaire}
\author[19]{Terrence Bohlsen}
\author[20]{Michel Pujol}
\author[21]{Etienne Bertrand}
\author[22]{Erik Bryssinck}
\author[23]{Val\'erie Desnoux}
\author[24]{Patrick Lailly}
\author[25]{Jacques Montier}
\author[26]{Massimiliano Mannucci}
\author[26]{Nico Montigiani}
\author[27]{Albert Stiewing}
\author[28]{James Daglen}
\author[29]{Christian Kreider}
\author[30]{Thierry Lemoult}
\author[31]{Tony Rodda}

\address[1]{Groupe d'Astrophysique des Hautes Energies, STAR, Universit\'e de Li\`ege, Quartier Agora (B5c, Institut d'Astrophysique et de G\'eophysique), All\'ee du 6 Ao\^ut 19c, B-4000 Sart Tilman, Li\`ege, Belgium}
\address[2]{Balmes 2, 08784 Piera (Barcelona), Spain}
\address[3]{Gooiergracht 91, 1250--1252 Laren, Netherlands}
\address[4]{Observatoire de la Tourbi\`ere, 38690 Chabons, France}
\address[5]{Observatoire de la Belle \'Etoile, 38420 Revel, France}
\address[6]{Verny Observatory, 57420 Verny, France}
\address[7]{Emil-Nolde-Strasse 12, 51375 Leverkusen, Germany}
\address[8]{PO box 141, Alpine, Texas 79831, USA}
\address[9]{Cotoletta Observatory, Via Matteotti 81, 20064 Gorgonzola (MI), Italy}
\address[10]{Kapellenkamp 21, 23569 L\"ubeck, Germany}
\address[11]{Uhlandstrasse 11d, 23617 Stockelsdorf, Germany}
\address[12]{Bellavista Observatory, Via Carlo De Paulis, 15 - 67100 L'Aquila, Italy}
\address[13]{1178 Mill Ridge Road, Arnprior, ON, K7S3G8, Canada}
\address[14]{39 Rue des Songes , 68850 Staffelfelden, France}
\address[15]{19 Bd Carnot, 21000 Dijon, France}
\address[16]{Observatoire du Pilat, 42660 Tarentaise, France}
\address[17]{23746 Schoolhouse Road, Manhattan, Illinois, USA}
\address[18]{Observatoire du Val de l'Arc, 13530 Trets, France}
\address[19]{Mirranook Observatory, Armidale NSW, Australia}
\address[20]{Rue des Pins, 31700 Beauzelle, France}
\address[21]{105 Bd de la C\^ote de Beaut\'e, 17640 Vaux Sur Mer, France}
\address[22]{Eyckensbeekstraat 2, 9150 Kruibeke, Belgium}
\address[23]{13 Rue Saint Charles, 75015 Paris, France}
\address[24]{Observatoire Canigou, 64000 Pau, France}
\address[25]{Centre d'Astronomie de La Couy\`ere, 35320 La Couy\`ere, France}
\address[26]{Osservatorio Astronomico Margherita Hack, Lasra a Signa (FI), Italy}
\address[27]{16210 N Desert Holly Dr, Sun City, AZ 85351, USA}
\address[28]{33 Joy Rd, PO Box 196,  Mayhill, New Mexico 88339, USA}
\address[29]{17 Rue Schelbaum, 68360 Soultz, France}
\address[30]{Chelles Observatory, 23 Avenue Henin, 77500 Chelles, France}
\address[31]{1 Rivermede, Ponteland, Newcastle upon Tyne NE20 9XA, UK}
  
\begin{abstract}
Some Be stars display important variability of the strength of the emission lines formed in their disk. This is notably the case of $\pi$\,Aqr. We present here the recent evolution of the Be disk in this system thanks to spectra collected by amateur spectroscopists since the end of 2013. A large transition occurred: the emission linked to the Be disk nearly disappeared in January 2014, but the disk has recovered, with a line strength now reaching levels only seen during the active phase of 1950--1990. In parallel to this change in strength occurs a change of disk structure, notably involving the disappearance of the strong asymmetry responsible for the $V/R$ modulation.
\end{abstract}


\begin{highlights}
\item The disk surrounding the primary component of $\pi$\,Aqr\ nearly disappeared in early 2014.
\item The disk has slowly recovered, now reaching strengths not seen in three decades.
\item This evolution in line strength is accompanied by changes in disk structure.
\end{highlights}


\end{frontmatter}

\section{Introduction}
$\pi$\,Aqr (HD212571) is a nearby ($d$=257--331\,pc, \citealt{bai18}) binary system comprising a Be primary \citep{bjo02}. Because of its brightness ($V$=4.64), it has been studied for a long time, both in photometry and spectroscopy. However, it was only recently found to belong to the peculiar class of $\gamma$-Cas stars, which gathers Be stars with unusually bright and hard X-ray emission \citep[see also \citealt{smi16} for a detailed review on such objects]{naz17}.

Variability is commonly seen in Be stars in general and in $\pi$\,Aqr in particular. After an active phase in 1950--1990, $\pi$\,Aqr went through a very weak emission phase in 1996--2000. \citet{bjo02} extensively studied it, which allowed them to detect the individual signatures of the two components of the system. As the disk had (nearly) disappeared, the H$\alpha$ line appeared mostly in absorption. This absorption component belonged to the massive primary. Superimposed on it were weak emissions due to the tenuous remnants of the disk and another weak emission displaying opposite motion with respect to the absorption. This emission was therefore attributed to the secondary star of the system. \citet{bjo02} measured the radial velocities ($RVs$) of the absorption and the secondary emission, and used them to derive the system's properties. $\pi$\,Aqr has an orbital period of 84.1d and a mass ratio of 0.16 which suggests a 2--3\,M$_{\odot}$ secondary star to the B1 primary.

\citet{zha13} further examined the variations of the H$\alpha$ line between 2004 and 2013. At that time, the activity oscillated between low and moderate states, without reaching the high levels of the active phase. Using tomographic techniques, \citet{zha13} showed that the disk was influenced by the presence of the companion, with the brightest regions located on the outer part of the disk facing the secondary.

As the star continues to evolve, we examine in this contribution the behaviour of $\pi$\,Aqr over the last six years. The observations used to this aim and their treatment are presented in the next section, while results are presented in detail in Sect. 3, with a summary ending the paper.

\section{Observations and data reduction}
Because of their brightness and varying character, Be stars are regularly observed by the amateur astronomer community. Such observations may be photometric or spectroscopic, in which case they usually cover the H$\alpha$ line. The Be Star Spectra (BeSS) open-access database\footnote{http://basebe.obspm.fr} \citep{nei11} collects and centralizes the spectra of Be stars taken by amateurs and professionals to ensure their legacy to the astronomical community as a whole. We have downloaded from it all spectra of $\pi$\,Aqr taken between Oct. 2013 and Jan. 2019 (i.e. since the analysis of \citealt{zha13}). All these spectra have been taken by amateurs, co-authors of this paper, who reduced the data in a standard way. More information on these data can be found on the BeSS website.

In addition to BeSS data, two German amateurs (C.T. Quandt and M. Moll) provided additional spectra of $\pi$\,Aqr in 2017--2018. They were taken using a 5 inch Schmidt-Cassegrain telescope equipped with a LHiRes III spectrograph. The detector was a Sony IMX 674 CCD in 2017 and a Sony IMX 694 CCD in 2018. Several exposures were taken and combined to get the final 13 spectra. Data reduction was carried out in ESO-MIDAS in a standard way. For wavelength calibration of the spectra, the internal NeAr lamp of the spectrograph was used but recalibration, performed using telluric lines, enabled them to improve the wavelength accuracy (on average, the precision is 3\,km\,s$^{-1}$).  

In total, our dataset consists of 379 H$\alpha$ spectra of $\pi$\,Aqr. The annual visibility season typically extends from June to January of the following year, but two-thirds of spectra were actually taken in October--December. Because we gather the contributions of different observers from various places, the spectra were taken in various meteorological and instrumental conditions. In particular, the used telescopes had diameters between 5 and 20 inches, spectrographs had resolving powers between 5000 and 20\,000 (with the exception of two low-resolution spectra taken at $R=600$ on 2015-11-08 by C. Kreider and 2018-08-22 by T. Rodda). Exposure times ranged from 200s to 7800s and the vast majority of spectra are of good quality (SNR of at least 100 in 87\% of spectra).

While the amateurs' spectra were roughly normalized, we have further processed them, first by correcting for telluric absorptions within IRAF using the template of \citet{hin00} then by applying a final normalization using continuum windows and a low-order polynomial. The continuum windows were adapted to the spectral range: in some cases, the range covered by the spectra was rather small hence the continuum windows had to be taken close to the line and the normalization was then poorer when the line was stronger and broader, which was the case in recent years (this explains some additional scatter of measurements for the last observing year). 

\begin{figure}
\begin{center}
\includegraphics*[width=9cm]{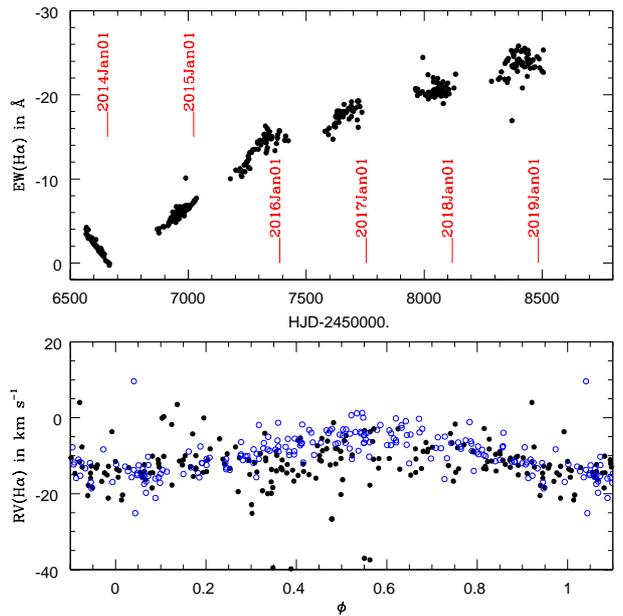}
\end{center}
\caption{{\it Top:} Evolution of H$\alpha$ $EWs$ in $\pi$\,Aqr since October 2013. {\it Bottom:} Measured $RVs$ (Oct. 2013--Feb. 2016 shown in black dots and July 2016--Jan. 2019 in blue circles), phased with average ephemeris from \citet[i.e. $P$=84.1\,d and $HJD_0$=2\,450\,275.5]{bjo02}. \label{ew}}
\end{figure}

\section{Results}
After applying the individual heliocentric corrections\footnote{Note that the keyword BSS\_RQVH in the BeSS header provides the opposite of that correction.}, we measured the moments of the H$\alpha$ line in all normalized spectra over the velocity interval --540\,km\,s$^{-1}$ to 540\,km\,s$^{-1}$ (using a rest wavelength of 6562.85\,\AA). No correction for the underlying photospheric absorption was performed, as in \citet{zha13} but contrary to \citet{bjo02}. The resulting moments are listed in the fifth to seventh columns of Table \ref{res}. The zeroth-order moment ($M_0=\sum (F_i-1)$) corresponds to the equivalent width ($EW$), i.e. it provides the width of a rectangular line of unity amplitude with the same integrated area as the observed line. Note that, in this paper, emission lines have negative EWs while absorptions have positive EWs. The first-order moment ($M_1=\sum (F_i-1)\times v_i /\sum (F_i-1)$) provides a flux-weighted centroid for the line, i.e. its $RV$. The second-order moment ($M_2=\sum (F_i-1)\times (v_i-M_1)^2 /\sum (F_i-1)$) provides the square of the line width; it would be equal to $\sigma^2$ if the line were a centered Gaussian. In parallel, we have also measured the height above continuum of the blue and red peaks and Table \ref{res} provides their ratio (traditionally called $V/R$). Comparing our results with \citet{zha13}, who did not use moments to derive $EWs$, we found that their $EWs$ and $V/R$ values for the data in October 2013 agree well with ours. We also compared the results for different observers (those who provided at least 20 spectra) and found no systematic difference between them. As in \citet{zha13}, we do not provide formal errors (e.g. from error propagation) on our measurements as the scatter of values are best representative of the actual errors, due to noise but also to normalization errors, imperfect telluric corrections, wavelength calibration errors,... We now examine each diagnostic in turn.

\subsection{EWs}

Figure \ref{ew} shows the evolution of the $EW$ over recent years. As can be seen, it decreases at the end of 2013 and then reaches $\sim$0\,\AA\ in January 2014. At that time, the emission had not completely disappeared, however. Rather, the emission component was so weak that the absorption component was revealed. Its strength nearly compensated that of the emission, leading to a $\sim$0\,\AA\ $EW$ value for the overall line. The line did not reach the absorption-dominated stage seen by \citet{bjo02}, but it was without doubt in a very low emission state. Such (nearly) disappearance of the emission is not infrequent in $\pi$\,Aqr. For example, \citet{mcl62} reported an absence of bright emission lines in 1936-7, 1944-5, and 1950 while \citet{zha13} reported another low emission state ($|EW|<1$\,\AA) from mid-2006 to mid-2007.

\begin{figure}
\begin{center}
\includegraphics*[width=9cm,bb=19 392 592 684, clip]{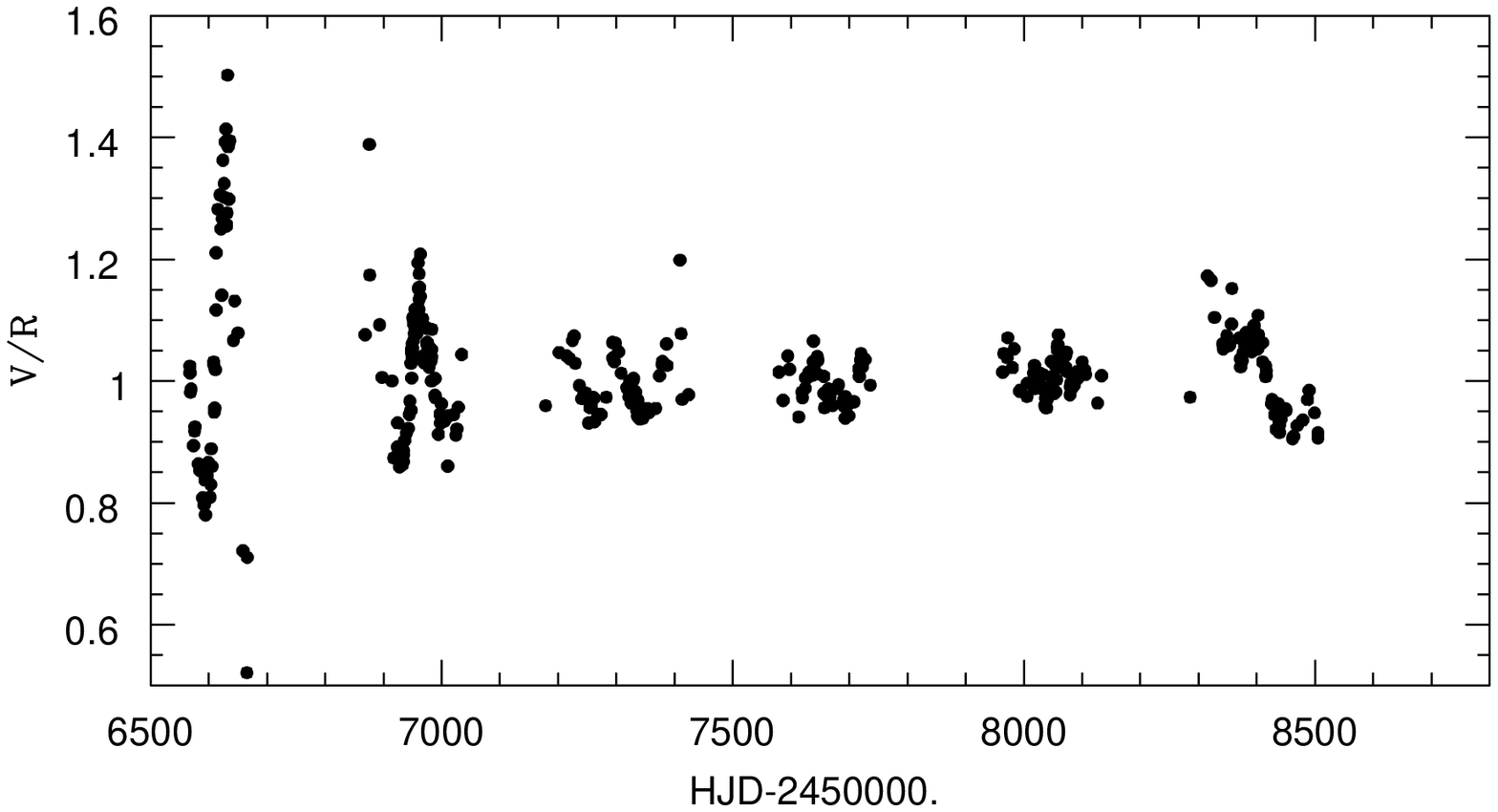}
\includegraphics*[width=9cm]{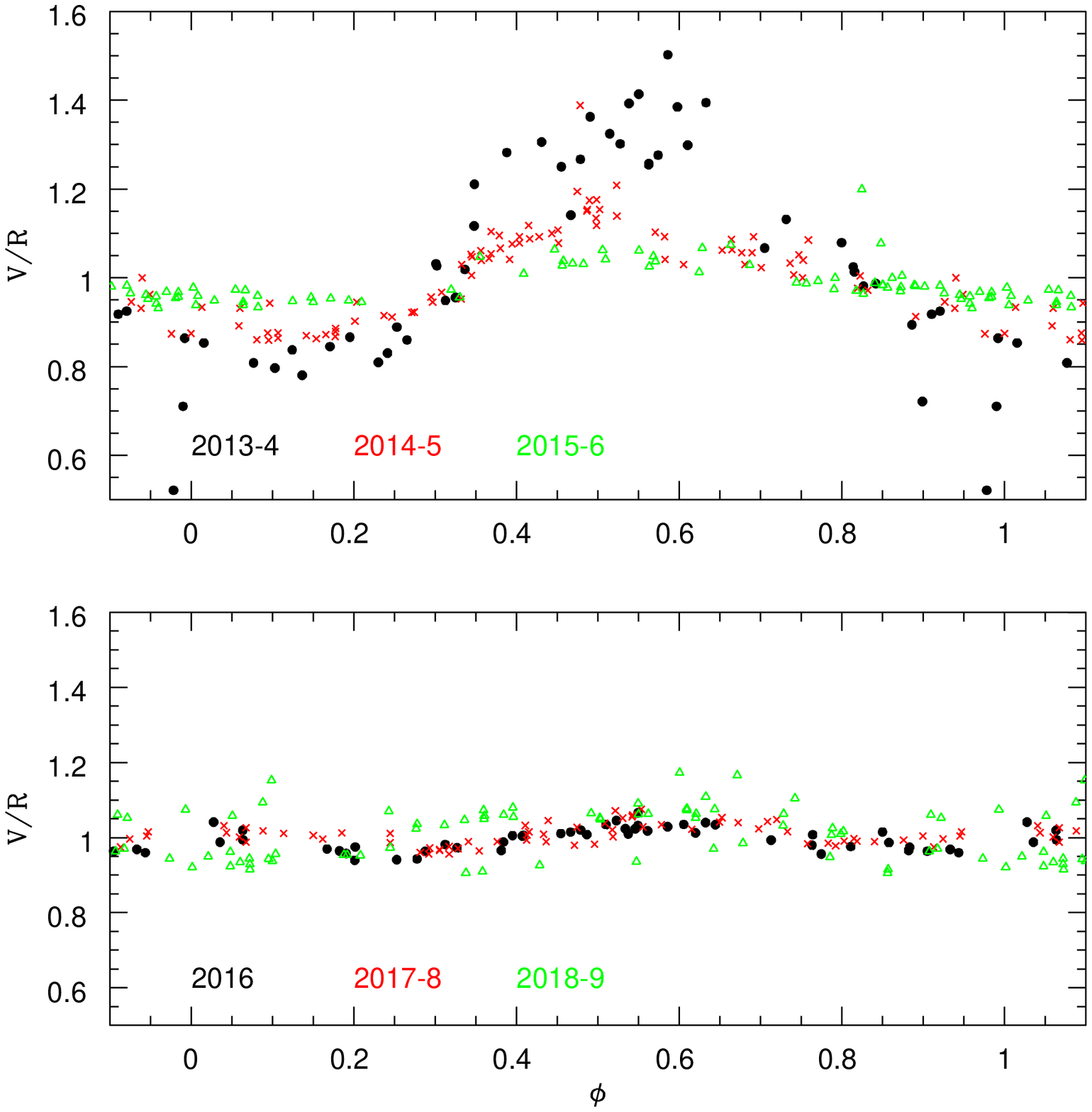}
\end{center}
\caption{Evolution of $V/R$ with time (top panel) or phase (middle and bottom panels, using the same ephemeris as for the RVs in Fig. \ref{ew}). In the lower panels, different colors and symbols are used to identify the observing seasons: black dots for 2013-4, red crosses for 2014-5, and empty green triangles for 2015-6 in the middle panel; black dots for 2016, red crosses for 2017-8, and empty green triangles for 2018-9 in the bottom panel. \label{vr}}
\end{figure}

After that minimum, the emission monotonically increased over the years, reaching $EW$ values around --25\,\AA\ in the last year. This represents a large change compared to recent years. Indeed, between 2004 and 2013, the H$\alpha$ line displayed at most a moderate emission, with absolute $EW$ values varying between 1 and 11\,\AA\ \citep{zha13}. Starting in Fall 2015, the $EWs$ exceeded such values, more than doubling in recent years. Figure 1 in \citet{bjo02} shows that large $EW$ values (--20 to --40\,\AA, after correction for photospheric absorption) were previously observed between 1950 and 1990. $\pi$\,Aqr thus seems to have entered a new active phase, with the low activity state of previous years coming to an end.

\subsection{RV}
Since $\pi$\,Aqr is a binary, its lines should regularly shift with orbital phase. Figure \ref{ew} shows the velocities of the H$\alpha$ line, phased with the ephemeris of \citet{bjo02}. While the scatter is not negligible, a clear sinusoidal variation, independent of the observing season, is detected, as could be expected. 

\subsection{$V/R$}
When in emission, the H$\alpha$ line of $\pi$\,Aqr usually appears double-peaked. This is a common feature in Be stars seen under a high inclination. In a significant fraction of such stars, monitorings revealed variations of the amplitude of the violet peak with respect to that of the red one \citep[see e.g.][]{por03}. In $\pi$\,Aqr, their ratio, called $V/R$, was found to undergo a sinusoidal variation with the same periodicity as the orbital motions \citep{zha13}. This modulation was attributed to the disk asymmetry triggered by the presence of the companion \citep{zha13}.

Figure \ref{vr} shows the evolution of the $V/R$ ratio with time and orbital phase. It clearly appears that the $V/R$ modulation depends on the line strength: the largest amplitudes in $V/R$ occur when the line is weakest. However, this is not a simple dilution effect: in the most recent data, the $V/R$ modulation actually disappears, with $V/R$ ratios simply scattered over phase. 

\subsection{Line profile}
\begin{figure}
\begin{center}
\includegraphics*[width=9cm]{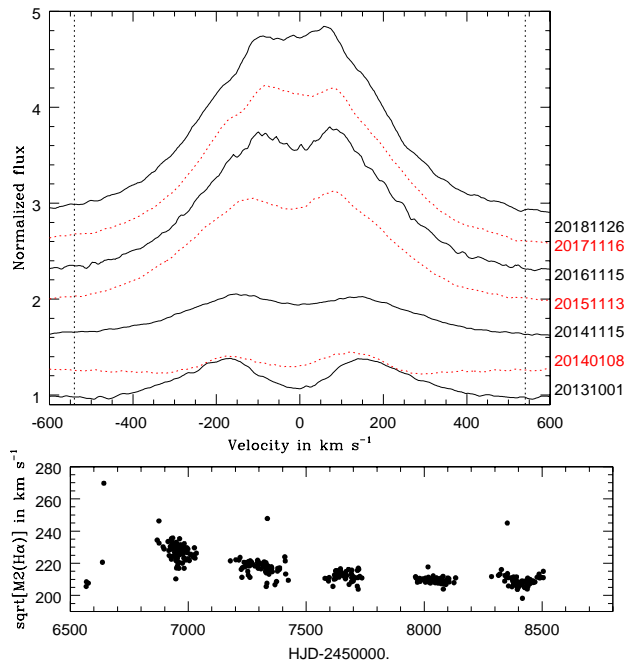}
\end{center}
\caption{Evolution of the H$\alpha$ line profile and its width (as measured by the second-order moment) since Oct. 2013. In the top panel, the individual spectra are shifted by a constant step of 0.3 to facilitate the comparison and velocities are heliocentric. \label{prof}}
\end{figure}

The top panel of Fig. \ref{prof} shows the evolution of the line profile variations with time. One profile per observing season is shown, except for 2013-4 for which the profiles at the beginning and end of the season are displayed. The double-peaked nature of the profile remained over the years. We have not yet reached a state with a single-peaked profile \citep{sle78} but the situation appears close to the asymmetric profile with two barely distinguishable peaks reported by \citet{and82}.

However, large changes in the profiles are detected besides the obvious variations in strength. First, the width of the line profile has continuously decreased as the line strength increased. This can be best seen in the bottom panel of Fig. \ref{prof}, where the second-order moment\footnote{In 2013-4, the line profile is a mix of absorption and emission with similar strengths, hence the first and second-order moments are not fully representative of the actual centroid and width of the emission.} evolves from $\sim$230 to $\sim$210\,km\,s$^{-1}$ between Fall 2014 and Winter 2018-9. This is in line with \citet{mcl62}, who already noted that reappearing Hydrogen emissions are broader than usual, leading to a width decrease afterwards. Second, at the same time, the two peaks clearly move closer to each other, with a separation changing from $\sim$320 to $\sim$140\,km\,s$^{-1}$. If linked to Keplerian rotation of the disk, this suggests an increase in the radial extension of the disk \citep{hum95}. The observed separation decrease translates into an increase in the radius of the H$\alpha$ emission region by a factor of $\sim$5 \citep[see their eq. 2]{zam19}.

The change in behaviour of the $V/R$ ratios and in the line profile can be taken as indicating a change in the disk structure. To further examine this transition, we have performed a tomographic analysis of the profiles in each season. This technique assumes that the emitting gas is stationary in the rotating frame of the binary, i.e. that the line profile variations are only due to our changing viewing angle because of the orbital motion \citep{hor91}. In this case, each emitting parcel is associated to a specific $(v_x,v_y)$ pair, where the x-axis points from the primary to the secondary and the y-axis has the same direction as the velocity vector of the secondary. The radial velocity of the emission, as seen by an observer on Earth, is then given at any phase by $v(\phi_t)=-v_x \cos(2\pi\phi_t)+v_y\sin(2\pi\phi_t)+v_z$ where $\phi_t$ is zero at the conjunction with the secondary star in front\footnote{Therefore $\phi_t=\phi+0.25$, where the orbital phase $\phi$ used in the rest of the paper refers to the ephemeris of \citet{bjo02}.}. We used an implementation of Doppler tomography relying on a Fourier-filtered back-projection algorithm, as in \citet{rau02,rau05}.

\begin{figure*}
\begin{center}
\includegraphics*[width=19cm, bb= 67 211 575 335, clip]{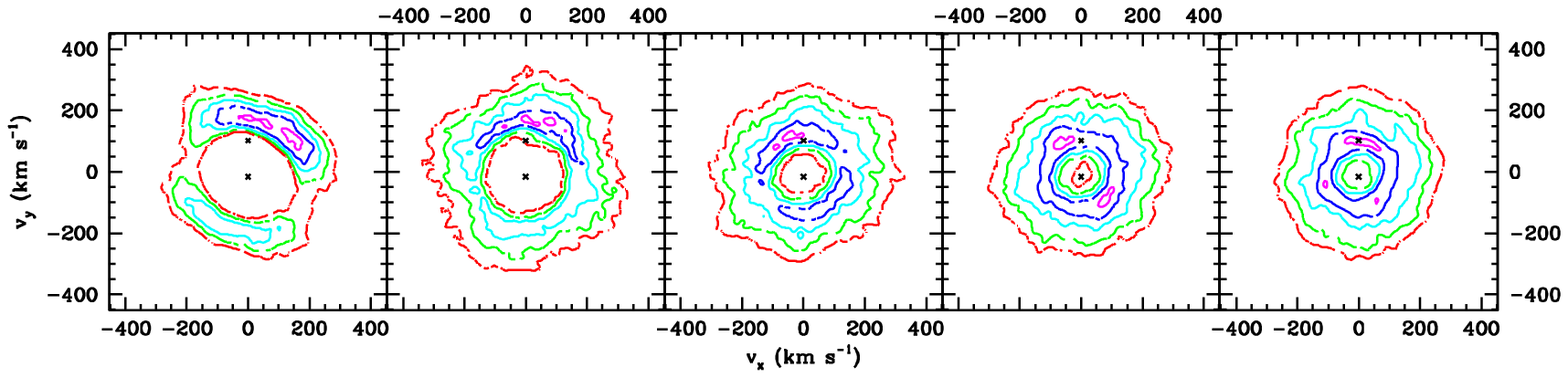}
\includegraphics*[width=16.5cm, bb=1 1 1220 240]{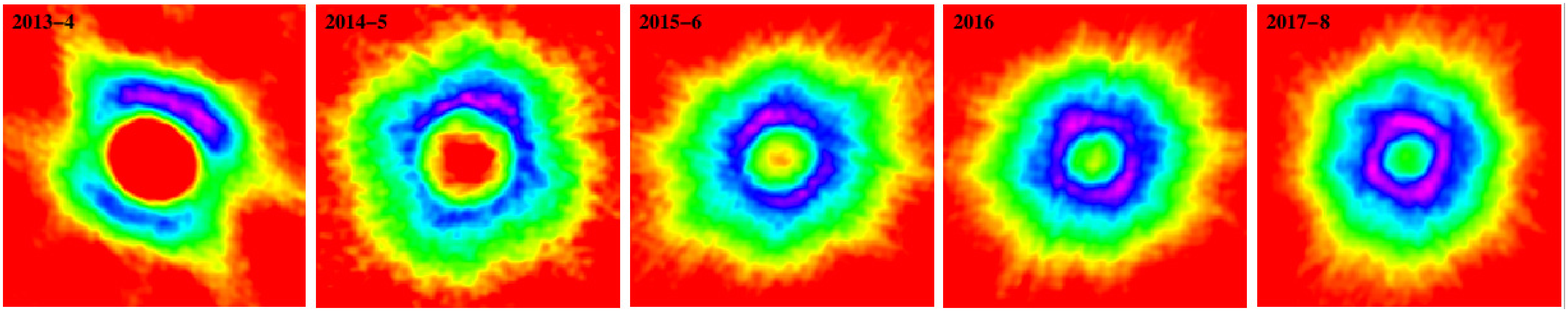}
\end{center}
\caption{Doppler maps for H$\alpha$ in the five observing seasons 2013-4, 2014-5, 2015-6, 2016, 2017-8. In the contour plots, the magenta, blue, cyan, green, and red contours correspond to amplitudes of 95\%, 80\%, 65\%, 50\%, and 35\% of the maximum emission. The two crosses indicate the velocities of the secondary (top) and primary (bottom) according to the semi-amplitudes $K$ derived by \citet{bjo02}. The bottom color images provide the same maps as images. These maps correspond to a slice at $v_z=-10$\,km\,s$^{-1}$, the mean $RV$ value of the system (see Fig. \ref{ew}) - calculations made with values of +10 or 0\,km\,s$^{-1}$ provide similar results, however. \label{tomo}}
\end{figure*}

To perform this analysis, we have separated the spectra by observing season, excluding the two taken at very low resolution ($R=600$) and those with $EW>-0.5$\,\AA\ (i.e. with a line strongly affected by absorption). Since the H$\alpha$ line strength significantly changes even during a single observing season, we have divided the individual line fluxes by the $EWs$. Furthermore, the spectra were weighted to achieve an orbital phase coverage as homogeneous as possible (i.e. if several spectra were taken at similar phases, each one had a reduced weight). Finally, the Doppler maps were calculated twice: first by considering all available spectra in each season and second by considering only a subset of spectra taken during that season by a single observer (e.g. J. Guarro-Fl\'o in 2013-4, A. De Bruin in 2016). The latter case provides more homogeneity (a single observer, a single observing place, a single instrument), which enables us to check whether combining data from several sources caused problems. We found that the maps were similar, but with lower SNR with fewer spectra, hence Fig. \ref{tomo} presents only the ones derived from all available spectra. We recall that they provide a measure of the line flux in the {\it velocity} space, which does not directly reflect {\it spatial} distribution. In particular, emission linked to a Keplerian disk should appear as an annulus with inner and outer regions inverted in velocity space - indeed, the inner disk regions have the largest velocities hence appear on the outer part of the Doppler map.

In the maps, the Be disk of $\pi$\,Aqr appears as a large structure of approximately annular form. The inner radius of this ring (central red contour in Fig. \ref{tomo}), the position of the maximum emission regions (blue and magenta contours in Fig. \ref{tomo}), as well as the (a)symmetric character of the overall structure appear to change with time. At first, the disk shows a strong asymmetry peaking on the secondary side, as found by \citet{zha13}. However, this asymmetry subsequently decreases and nearly disappears after 2015 or 2016. This may seem to contradict the previous results of \citet{zha13} who had found that the overall inhomogeneity in disk surface brightness was larger when the line was stronger, but we recall that the regime in line strength probed in that paper is very different than observed since Fall 2015. There is thus no contradiction, only a change in behaviour in the most active phase. The strong reduction in disk asymmetry in Doppler maps is in line with the disappearance of the $V/R$ modulation. At the same time, the maps indicate that the inner edge of the disk gets closer (in velocity) to the Be star, suggesting again a larger disk extension.

\section{Conclusion}
In this paper, we report on changes in the H$\alpha$ line profile of the Be star $\pi$\,Aqr since the last published monitoring \citep{zha13}. During Winter 2013-2014, the emission of $\pi$\,Aqr became very weak but did not disappear completely. Its strength was then similar to that of the photospheric absorption. Since then, the emission has monotonically increased and is now reaching levels only seen during the last ``active'' phase, in 1950-1990. In this process, the Be disk of $\pi$\,Aqr appears to have extended and to have lost the strong asymmetry responsible for the $V/R$ modulation of the line peaks. 

\begin{table*}
  \caption{Journal of the observations of $\pi$\,Aqr taken between Oct. 2013 and Jan. 2019, as well as the associated measurements of the H$\alpha$ line. As observing log are provided the observing date, observer ID (using the BeSS nickname when applicable, initials otherwise), and instrument details (telescope size and resolving power of spectrograph). The provided H$\alpha$ results are $EWs$ (0th-order moment), $RVs$ (1st-order moment), widths (squared root of the 2nd-order moment) and $V/R$ ratios. Note that, when the line combines absorption and emission (i.e. 0th-order moment is close to zero), the higher order moments become less representative of the properties of the emission component. Low-resolution data ($R$=600) also have imprecise higher order moments.}
\label{res}
\begin{tabular}{l c c c c c c c }
  \hline
  Date & $HJD$ & Observer & tel size/$R$ & $EW$ & $RV$  & width & $V/R$ \\
  (YYYYMMDD) & $-2\,450\,000$ & & (mm)/ & (\AA) & (km\,s$^{-1}$) & (km\,s$^{-1}$) &\\
\hline
20131001 &  6567.358 &   epollmann & 400/20000&   -3.99 &   -13.4 & 206 & 1.025 \\ 
20131001 &  6567.476 &  jguarroflo & 406/5000 &   -3.45 &    -2.8 & 188 & 1.013 \\ 
20131002 &  6568.427 &    fhoupert & 280/17000&   -4.23 &    -5.6 & 208 & 0.981 \\ 
20131004 &  6569.696 &    csawicki & 356/15000&   -3.38 &   -14.0 & 188 & 0.987 \\ 
20131007 &  6573.425 &  jguarroflo & 406/5000 &   -3.27 &   -10.4 & 181 & 0.894 \\ 
20131009 &  6575.461 &  jguarroflo & 406/5000 &   -2.95 &   -14.7 & 169 & 0.918 \\ 
20131010 &  6576.336 &   epollmann & 400/20000&   -3.97 &     4.0 & 208 & 0.925 \\ 
20131016 &  6582.332 &  jguarroflo & 406/5000 &   -3.03 &    -3.7 & 175 & 0.864 \\ 
20131018 &  6584.309 &  jguarroflo & 406/5000 &   -3.10 &   -20.3 & 170 & 0.853 \\ 
20131023 &  6589.455 &  jguarroflo & 406/5000 &   -2.71 &   -14.6 & 145 & 0.808 \\ 
20131026 &  6591.654 &    csawicki & 356/15000&   -2.99 &    -0.1 & 171 & 0.797 \\ 
20131027 &  6593.453 &  jguarroflo & 406/5000 &   -2.49 &    -1.8 & 131 & 0.837 \\ 
20131028 &  6594.472 &  jguarroflo & 406/5000 &   -2.65 &     3.5 & 155 & 0.780 \\ 
20131031 &  6597.365 &  jguarroflo & 406/5000 &   -2.52 &    -4.2 & 140 & 0.845 \\ 
20131102 &  6599.414 &  jguarroflo & 406/5000 &   -2.28 &    -0.1 & 114 & 0.866 \\ 
20131105 &  6602.348 &  jguarroflo & 406/5000 &   -2.21 &    -5.5 &  99 & 0.809 \\ 
20131106 &  6603.305 &  jguarroflo & 406/5000 &   -2.16 &   -12.2 & 100 & 0.830 \\ 
20131107 &  6604.280 &  jguarroflo & 406/5000 &   -2.29 &   -13.3 & 124 & 0.889 \\ 
20131108 &  6605.319 &  jguarroflo & 406/5000 &   -2.11 &    -7.7 &  91 & 0.860 \\ 
20131111 &  6608.309 &  jguarroflo & 406/5000 &   -2.01 &   -22.9 &  74 & 1.031 \\ 
20131111 &  6608.406 &  bmauclaire & 300/16000&   -2.79 &   -25.2 & 125 & 1.027 \\ 
20131112 &  6609.298 &  jguarroflo & 406/5000 &   -1.92 &   -14.3 &   6 & 0.949 \\ 
20131113 &  6610.332 &  jguarroflo & 406/5000 &   -1.98 &   -19.2 &  78 & 0.955 \\ 
20131114 &  6611.311 &  jguarroflo & 406/5000 &   -2.01 &   -20.0 &  98 & 1.019 \\ 
20131115 &  6612.261 &  jguarroflo & 406/5000 &   -1.97 &   -18.4 &  75 & 1.117 \\ 
20131115 &  6612.290 &   epollmann & 400/20000&   -1.97 &   -39.5 &  87 & 1.211 \\ 
20131119 &  6615.636 &    csawicki & 356/15000&   -1.73 &   -39.8 &  84 & 1.282 \\ 
20131122 &  6619.242 &  jguarroflo & 406/5000 &   -1.69 &   -45.3 &  59 & 1.306 \\ 
20131124 &  6621.280 &  jguarroflo & 406/5000 &   -1.40 &   -11.4 & 137 & 1.250 \\ 
20131125 &  6622.269 &  jguarroflo & 406/5000 &   -1.71 &    -8.2 &  11 & 1.141 \\ 
20131126 &  6623.251 &  jguarroflo & 406/5000 &   -1.47 &   -26.6 & 118 & 1.267 \\ 
20131127 &  6624.254 &  jguarroflo & 406/5000 &   -1.47 &   -59.4 & 117 & 1.363 \\ 
20131129 &  6626.283 &  jguarroflo & 406/5000 &   -1.41 &   -40.8 & 122 & 1.324 \\ 
20131130 &  6627.358 &  jguarroflo & 406/5000 &   -1.46 &   -54.4 & 105 & 1.302 \\ 
20131201 &  6628.270 &  jguarroflo & 406/5000 &   -1.31 &   -42.0 & 137 & 1.393 \\ 
20131202 &  6629.276 &   epollmann & 400/20000&   -1.75 &   -37.0 &  68 & 1.414 \\ 
20131203 &  6630.311 &  jguarroflo & 406/5000 &   -1.11 &   -17.8 & 177 & 1.255 \\ 
20131203 &  6630.318 &  bmauclaire & 300/16000&   -1.64 &   -37.4 &  30 & 1.257 \\ 
20131204 &  6631.284 &  jguarroflo & 406/5000 &   -1.36 &   -42.4 &  81 & 1.276 \\ 
20131205 &  6632.287 &  jguarroflo & 406/5000 &   -1.12 &   -56.2 & 178 & 1.503 \\ 
20131206 &  6633.274 &  jguarroflo & 406/5000 &   -1.16 &   -41.3 & 162 & 1.385 \\ 
20131207 &  6634.345 &  bmauclaire & 300/16000&   -1.43 &   -41.0 &  80 & 1.299 \\ 
20131209 &  6636.237 &  jguarroflo & 406/5000 &   -0.94 &   -64.5 & 221 & 1.395 \\ 
20131215 &  6642.293 &  jguarroflo & 406/5000 &   -0.76 &   -63.3 & 270 & 1.067 \\ 
20131218 &  6644.525 &     kgraham & 305/13500&   -1.06 &   -45.3 & 171 & 1.131 \\ 
20131223 &  6650.248 &  jguarroflo & 406/5000 &   -0.28 &  -161.7 & 565 & 1.079 \\ 
20140101 &  6658.605 &    csawicki & 356/15000&   -0.12 &   -69.6 & 854 & 0.721 \\ 
20140107 &  6665.266 &  jguarroflo & 406/5000 &    0.27 &   -90.3 & 643 & 0.521 \\ 
20140108 &  6666.261 &  jguarroflo & 406/5000 &    0.04 &   -69.2 &1479 & 0.710 \\ 
20140729 &  6868.478 &    vdesnoux & 200/12000&   -4.02 &   -16.7 & 234 & 1.076 \\ 
20140805 &  6875.509 &   epollmann & 400/20000&   -3.59 &   -26.7 & 246 & 1.389 \\ 
\hline
\end{tabular}
\end{table*}
\setcounter{table}{0}
\begin{table*}
\caption{Continued.}
\begin{tabular}{l c c c c c c c }
  \hline
  Date & $HJD$ & Observer & tel size/$R$ & $EW$ & $RV$  & width & $V/R$ \\
  (YYYYMMDD) & $-2\,450\,000$ & & (mm)/ & (\AA) & (km\,s$^{-1}$) & (km\,s$^{-1}$) &\\
\hline
20140806 &  6876.470 &  bmauclaire & 300/16000&   -4.09 &    -8.2 & 232 & 1.174 \\ 
20140823 &  6893.470 &   epollmann & 400/20000&   -4.33 &    -8.3 & 230 & 1.092 \\ 
20140828 &  6897.605 &  bmauclaire & 300/16000&   -4.30 &    -4.2 & 229 & 1.006 \\ 
20140913 &  6914.346 &   epollmann & 400/20000&   -4.46 &   -17.8 & 233 & 1.000 \\ 
20140916 &  6917.375 &  pfosanelli & 280/17000&   -4.84 &   -14.5 & 228 & 0.874 \\ 
20140918 &  6919.400 &  pfosanelli & 280/17000&   -4.71 &   -11.5 & 225 & 0.875 \\ 
20140923 &  6924.339 &   epollmann & 400/20000&   -5.13 &   -15.6 & 224 & 0.892 \\ 
20140923 &  6924.423 &  pfosanelli & 280/17000&   -5.19 &   -15.0 & 230 & 0.931 \\ 
20140926 &  6927.332 &    pberardi & 235/13500&   -5.34 &   -11.1 & 231 & 0.876 \\ 
20140926 &  6927.418 &  pfosanelli & 280/17000&   -5.09 &   -12.1 & 229 & 0.859 \\ 
20140927 &  6928.373 &  jguarroflo & 406/5000 &   -5.40 &   -10.2 & 235 & 0.864 \\ 
20140927 &  6928.375 &     plailly & 203/15000&   -5.35 &     0.3 & 223 & 0.876 \\ 
20140930 &  6931.308 &    pberardi & 235/13500&   -5.30 &   -11.4 & 229 & 0.869 \\ 
20141001 &  6932.344 &  jguarroflo & 406/5000 &   -5.30 &   -10.0 & 232 & 0.863 \\ 
20141002 &  6933.332 &  jguarroflo & 406/5000 &   -5.42 &   -12.9 & 233 & 0.872 \\ 
20141003 &  6934.303 &   epollmann & 400/20000&   -4.98 &   -13.6 & 224 & 0.868 \\ 
20141003 &  6934.333 &  jguarroflo & 406/5000 &   -5.33 &   -13.2 & 232 & 0.886 \\ 
20141003 &  6934.338 &    fhoupert & 280/15000&   -6.07 &   -10.0 & 236 & 0.878 \\ 
20141005 &  6936.333 &    pberardi & 235/13500&   -5.60 &   -14.2 & 230 & 0.902 \\ 
20141008 &  6939.364 &  jguarroflo & 406/5000 &   -5.66 &   -10.9 & 232 & 0.915 \\ 
20141011 &  6942.456 &  pfosanelli & 280/17000&   -5.48 &   -10.6 & 227 & 0.922 \\ 
20141013 &  6944.373 &  jguarroflo & 406/5000 &   -5.67 &   -14.9 & 230 & 0.945 \\ 
20141014 &  6945.305 &  pfosanelli & 280/17000&   -5.43 &   -11.8 & 224 & 0.967 \\ 
20141016 &  6947.316 &      mpujol & 300/11000&   -4.82 &   -10.6 & 210 & 0.952 \\ 
20141016 &  6947.358 &  jguarroflo & 406/5000 &   -5.67 &   -11.3 & 231 & 1.029 \\ 
20141017 &  6948.357 &  jguarroflo & 406/5000 &   -5.64 &   -16.9 & 227 & 1.052 \\ 
20141017 &  6948.393 &   mleonardi & 235/7800 &   -5.37 &   -19.9 & 222 & 1.005 \\ 
20141017 &  6948.421 &     plailly & 203/15000&   -5.89 &   -13.6 & 225 & 1.047 \\ 
20141018 &  6949.356 &   epollmann & 400/20000&   -5.49 &   -14.0 & 222 & 1.061 \\ 
20141018 &  6949.407 &    fhoupert & 280/15000&   -6.69 &    -9.9 & 233 & 1.040 \\ 
20141019 &  6950.398 &  pfosanelli & 280/17000&   -5.67 &   -11.1 & 224 & 1.054 \\ 
20141019 &  6950.409 &   mleonardi & 235/7800 &   -5.54 &   -13.7 & 217 & 1.104 \\ 
20141020 &  6951.274 &    pberardi & 235/13500&   -5.64 &   -14.5 & 224 & 1.095 \\ 
20141020 &  6951.359 &  jguarroflo & 406/5000 &   -5.84 &   -12.3 & 231 & 1.066 \\ 
20141021 &  6952.341 &  jguarroflo & 406/5000 &   -6.12 &   -13.8 & 231 & 1.042 \\ 
20141022 &  6953.321 &  jguarroflo & 406/5000 &   -5.86 &    -6.0 & 229 & 1.078 \\ 
20141022 &  6953.359 &   mleonardi & 235/7800 &   -5.82 &   -15.2 & 223 & 1.093 \\ 
20141023 &  6954.299 &      mpujol & 300/11000&   -5.38 &   -12.2 & 218 & 1.118 \\ 
20141023 &  6954.389 &   mleonardi & 235/7800 &   -5.74 &   -14.2 & 221 & 1.087 \\ 
20141024 &  6955.387 &   mleonardi & 235/7800 &   -5.42 &   -15.0 & 221 & 1.093 \\ 
20141026 &  6956.673 &    csawicki & 356/15000&   -6.35 &   -15.9 & 230 & 1.100 \\ 
20141026 &  6957.349 &   mleonardi & 235/7800 &   -6.19 &   -10.0 & 232 & 1.107 \\ 
20141026 &  6957.386 &  pfosanelli & 280/17000&   -5.79 &   -11.1 & 225 & 1.078 \\ 
20141028 &  6959.329 &   mleonardi & 235/7800 &   -6.03 &   -12.4 & 227 & 1.194 \\ 
20141029 &  6960.308 &      mpujol & 300/11000&   -5.33 &    -8.0 & 217 & 1.151 \\ 
20141029 &  6960.350 & nmont+mmann & 250/17000&   -5.24 &   -12.2 & 228 & 1.153 \\ 
20141030 &  6961.270 &      mpujol & 300/11000&   -5.52 &    -8.9 & 220 & 1.134 \\ 
20141030 &  6961.317 & nmont+mmann & 250/17000&   -5.66 &   -20.2 & 225 & 1.176 \\ 
20141030 &  6961.336 &   mleonardi & 235/7800 &   -6.04 &   -10.7 & 225 & 1.118 \\ 
20141031 &  6961.649 &    csawicki & 356/15000&   -6.22 &   -16.3 & 230 & 1.154 \\ 
20141101 &  6963.397 &   mleonardi & 235/7800 &   -6.19 &   -10.0 & 229 & 1.208 \\ 
20141101 &  6963.427 &    fhoupert & 280/15000&   -6.69 &    -9.8 & 235 & 1.139 \\ 
20141105 &  6967.388 &  jguarroflo & 406/5000 &   -6.19 &   -10.9 & 231 & 1.102 \\ 
20141106 &  6968.332 &  jguarroflo & 406/5000 &   -5.97 &   -13.3 & 230 & 1.093 \\ 
\hline
\end{tabular}
\end{table*}
\setcounter{table}{0}
\begin{table*}
\caption{Continued.}
\begin{tabular}{l c c c c c c c}
  \hline
  Date & $HJD$ & Observer & tel size/$R$ & $EW$ & $RV$  & width & $V/R$ \\
  (YYYYMMDD) & $-2\,450\,000$ & & (mm)/ & (\AA) & (km\,s$^{-1}$) & (km\,s$^{-1}$) &\\
\hline
20141106 &  6968.386 &   mleonardi & 235/7800 &   -6.15 &   -10.8 & 231 & 1.042 \\ 
20141108 &  6970.324 &  jguarroflo & 406/5000 &   -5.96 &   -10.7 & 229 & 1.030 \\ 
20141112 &  6974.323 &  jguarroflo & 406/5000 &   -5.95 &   -15.1 & 228 & 1.062 \\ 
20141113 &  6975.245 &   epollmann & 400/20000&   -5.94 &   -14.4 & 220 & 1.087 \\ 
20141113 &  6975.303 &   mleonardi & 235/7800 &   -6.33 &   -10.9 & 229 & 1.063 \\ 
20141114 &  6976.315 &    pberardi & 235/13500&   -5.98 &   -13.4 & 226 & 1.057 \\ 
20141115 &  6976.641 &    csawicki & 356/15000&   -6.59 &    -8.8 & 231 & 1.029 \\ 
20141115 &  6977.390 &  jguarroflo & 406/5000 &   -5.91 &    -9.0 & 228 & 1.057 \\ 
20141116 &  6978.369 &  jguarroflo & 406/5000 &   -6.19 &    -7.7 & 230 & 1.023 \\ 
20141119 &  6981.335 &   mleonardi & 235/7800 &   -6.72 &   -12.6 & 230 & 1.033 \\ 
20141120 &  6982.252 &    pberardi & 235/13500&   -5.97 &   -16.7 & 223 & 1.052 \\ 
20141121 &  6982.596 &     kgraham & 305/13500&   -5.60 &    -1.7 & 217 & 1.000 \\ 
20141121 &  6982.691 &    csawicki & 356/15000&   -6.08 &   -14.4 & 223 & 1.040 \\ 
20141121 &  6983.223 &   epollmann & 400/20000&   -6.57 &   -13.4 & 232 & 1.085 \\ 
20141126 &  6988.328 &  jguarroflo & 406/5000 &   -6.24 &   -13.5 & 227 & 0.976 \\ 
20141127 &  6988.592 &    csawicki & 356/15000&   -6.82 &   -11.5 & 229 & 1.004 \\ 
20141127 &  6989.359 &   epollmann & 400/20000&  -10.11 &   -11.4 & 222 & 0.972 \\ 
20141202 &  6994.307 &  jguarroflo & 406/5000 &   -6.60 &   -15.0 & 230 & 0.913 \\ 
20141205 &  6997.325 &  jguarroflo & 406/5000 &   -6.54 &   -13.0 & 227 & 0.946 \\ 
20141206 &  6998.345 &  jguarroflo & 406/5000 &   -6.85 &   -20.5 & 229 & 0.931 \\ 
20141207 &  6999.218 &    pberardi & 235/13500&   -6.40 &   -18.4 & 223 & 0.962 \\ 
20141213 &  7004.622 &    csawicki & 356/15000&   -6.53 &   -21.7 & 222 & 0.934 \\ 
20141218 &  7010.291 &   mleonardi & 235/7800 &   -6.88 &   -18.5 & 226 & 0.861 \\ 
20141220 &  7011.639 &    csawicki & 356/15000&   -6.90 &   -13.0 & 225 & 0.943 \\ 
20141229 &  7020.622 &    csawicki & 356/15000&   -7.13 &   -11.1 & 224 & 0.945 \\ 
20150101 &  7024.267 &   mleonardi & 235/7800 &   -7.30 &    -9.5 & 225 & 0.911 \\ 
20150103 &  7026.311 &   mleonardi & 235/7800 &   -7.44 &   -19.5 & 230 & 0.921 \\ 
20150105 &  7028.312 &   mleonardi & 235/7800 &   -7.41 &   -11.2 & 223 & 0.957 \\ 
20150111 &  7034.296 &   mleonardi & 235/7800 &   -7.70 &   -13.2 & 226 & 1.044 \\ 
20150605 &  7178.584 &      ogarde & 400/11000&  -10.02 &   -10.1 & 222 & 0.959 \\ 
20150628 &  7201.574 &    jmontier & 355/17000&  -11.05 &    -8.9 & 222 & 1.047 \\ 
20150711 &  7214.578 &   epollmann & 400/20000&  -10.98 &    -4.6 & 222 & 1.042 \\ 
20150716 &  7219.740 &     tlester & 310/8000 &  -11.22 &    -2.9 & 224 & 1.037 \\ 
20150721 &  7224.569 &      othizy & 280/10000&  -10.41 &   -13.5 & 216 & 1.067 \\ 
20150723 &  7227.487 &      jterry & 300/9000 &  -10.29 &    -8.2 & 211 & 1.074 \\ 
20150725 &  7229.453 &      ogarde & 400/11000&  -10.92 &    -6.5 & 217 & 1.029 \\ 
20150802 &  7236.520 &    jmontier & 355/17000&  -11.64 &    -7.1 & 221 & 0.992 \\ 
20150805 &  7240.463 &   epollmann & 400/20000&  -10.87 &   -12.8 & 218 & 0.971 \\ 
20150809 &  7243.737 &     tlester & 310/8000 &  -11.84 &   -11.2 & 220 & 0.978 \\ 
20150813 &  7247.539 &     afavaro & 200/17000&  -12.04 &   -10.5 & 219 & 0.980 \\ 
20150817 &  7252.425 &  bmauclaire & 300/16000&  -12.11 &   -12.4 & 215 & 0.931 \\ 
20150819 &  7254.462 &      jterry & 300/9000 &  -11.41 &   -21.2 & 213 & 0.957 \\ 
20150819 &  7254.488 &    adebruin & 280/5800 &  -12.69 &   -12.8 & 221 & 0.967 \\ 
20150821 &  7256.488 &     afavaro & 200/17000&  -12.37 &   -13.9 & 218 & 0.959 \\ 
20150825 &  7260.380 &      ogarde & 400/11000&  -13.06 &   -14.5 & 218 & 0.973 \\ 
20150826 &  7261.394 &      jterry & 300/9000 &  -11.23 &   -16.3 & 211 & 0.972 \\ 
20150828 &  7262.737 &     tlester & 310/8000 &  -13.13 &   -14.5 & 222 & 0.934 \\ 
20150902 &  7268.416 &    adebruin & 280/5800 &  -13.16 &   -11.1 & 221 & 0.945 \\ 
20150907 &  7273.412 &      ogarde & 400/11000&  -13.48 &    -8.2 & 219 & 0.945 \\ 
20150917 &  7282.710 &     tlester & 310/8000 &  -13.53 &    -9.4 & 218 & 0.973 \\ 
20150927 &  7293.370 &   epollmann & 400/20000&  -14.40 &    -7.4 & 221 & 1.063 \\ 
20150928 &  7294.297 &    adebruin & 280/5800 &  -14.29 &    -8.7 & 222 & 1.038 \\ 
20150930 &  7296.354 &    adebruin & 280/5800 &  -13.53 &    -1.3 & 217 & 1.032 \\ 
20151002 &  7298.362 &   epollmann & 400/20000&  -14.21 &    -6.1 & 219 & 1.062 \\ 
\hline
\end{tabular}
\end{table*}
\setcounter{table}{0}
\begin{table*}
\caption{Continued.}
\begin{tabular}{l c c c c c c c}
  \hline
  Date & $HJD$ & Observer & tel size/$R$ & $EW$ & $RV$  & width & $V/R$ \\
  (YYYYMMDD) & $-2\,450\,000$ & & (mm)/ & (\AA) & (km\,s$^{-1}$) & (km\,s$^{-1}$) &\\
\hline
20151008 &  7303.627 &     tlester & 310/8000 &  -14.61 &    -6.3 & 220 & 1.048 \\ 
20151012 &  7308.347 &    adebruin & 280/5800 &  -14.04 &    -2.9 & 217 & 1.013 \\ 
20151022 &  7318.415 &  jguarroflo & 406/5000 &  -14.60 &    -6.6 & 218 & 0.990 \\ 
20151023 &  7319.404 &  jguarroflo & 406/5000 &  -14.56 &    -7.9 & 219 & 0.987 \\ 
20151026 &  7322.296 &    adebruin & 280/5800 &  -15.31 &    -8.8 & 221 & 0.974 \\ 
20151026 &  7322.401 &   epollmann & 400/20000&  -15.25 &    -9.0 & 220 & 1.000 \\ 
20151029 &  7325.305 &    adebruin & 280/5800 &  -15.28 &    -8.2 & 220 & 0.964 \\ 
20151030 &  7326.495 &     tlester & 310/8000 &  -14.90 &    -9.7 & 218 & 0.988 \\ 
20151031 &  7327.267 &      ogarde & 400/11000&  -15.31 &    -8.9 & 217 & 0.984 \\ 
20151031 &  7327.349 &    fhoupert & 280/15000&  -16.30 &    -9.6 & 220 & 0.984 \\ 
20151101 &  7328.310 &    vdesnoux & 235/15000&  -14.32 &   -12.3 & 215 & 1.000 \\ 
20151102 &  7329.222 &    adebruin & 280/5800 &  -15.28 &   -13.1 & 220 & 0.979 \\ 
20151102 &  7329.296 &   epollmann & 400/20000&  -14.90 &   -10.6 & 218 & 1.004 \\ 
20151104 &  7330.585 &     kgraham & 305/13500&  -14.33 &   -10.1 & 214 & 0.984 \\ 
20151104 &  7330.704 &    csawicki & 510/17000&  -13.10 &   -12.5 & 206 & 0.981 \\ 
20151106 &  7333.245 &      ogarde & 400/11000&  -15.16 &   -11.6 & 217 & 0.982 \\ 
20151107 &  7333.650 &    csawicki & 510/17000&  -13.44 &    -7.7 & 208 & 0.965 \\ 
20151108 &  7335.244 &    pberardi & 235/13500&  -14.77 &   -14.6 & 216 & 0.961 \\ 
20151108 &  7335.278 &    ckreider & 430/600  &  -13.72 &   -15.9 & 248 & -     \\ 
20151108 &  7335.422 &    fhoupert & 280/15000&  -16.00 &   -12.7 & 217 & 0.952 \\ 
20151109 &  7336.310 &     afavaro & 200/17000&  -15.92 &   -13.2 & 218 & 0.958 \\ 
20151109 &  7336.320 &  jguarroflo & 406/5000 &  -14.90 &   -13.1 & 218 & 0.942 \\ 
20151110 &  7337.347 &  jguarroflo & 406/5000 &  -15.00 &   -16.8 & 218 & 0.968 \\ 
20151111 &  7338.335 &  jguarroflo & 406/5000 &  -14.83 &   -15.1 & 218 & 0.954 \\ 
20151113 &  7340.395 &  jguarroflo & 406/5000 &  -15.02 &   -15.7 & 217 & 0.938 \\ 
20151115 &  7342.329 &     plailly & 203/15000&  -15.05 &   -13.1 & 213 & 0.949 \\ 
20151118 &  7345.267 &    pberardi & 235/13500&  -15.06 &   -12.3 & 216 & 0.939 \\ 
20151118 &  7345.274 &     afavaro & 200/17000&  -15.63 &   -14.1 & 217 & 0.945 \\ 
20151123 &  7350.396 &  jguarroflo & 406/5000 &  -14.77 &   -17.7 & 217 & 0.948 \\ 
20151125 &  7352.334 &  jguarroflo & 406/5000 &  -14.79 &   -12.4 & 218 & 0.955 \\ 
20151127 &  7354.351 &  jguarroflo & 406/5000 &  -14.99 &   -15.5 & 217 & 0.954 \\ 
20151129 &  7356.217 &    pberardi & 235/13500&  -15.01 &   -14.4 & 215 & 0.949 \\ 
20151211 &  7367.673 &    csawicki & 510/17000&  -13.36 &    -4.1 & 207 & 0.955 \\ 
20151217 &  7374.283 &  pfosanelli & 280/16000&  -14.21 &    -5.3 & 209 & 1.008 \\ 
20151221 &  7378.271 &  jguarroflo & 406/5000 &  -14.84 &    -5.4 & 215 & 1.027 \\ 
20151222 &  7379.327 &  jguarroflo & 406/5000 &  -15.23 &    -6.1 & 217 & 1.032 \\ 
20151229 &  7386.227 &   mleonardi & 235/7800 &  -15.75 &    -3.4 & 216 & 1.061 \\ 
20151230 &  7387.265 &  jguarroflo & 406/5000 &  -15.67 &    -3.7 & 217 & 1.025 \\ 
20160121 &  7409.247 &  jguarroflo & 406/5000 &  -14.58 &   -71.0 & 224 & 1.199 \\ 
20160123 &  7411.243 &  jguarroflo & 406/5000 &  -14.62 &   -57.9 & 221 & 1.078 \\ 
20160125 &  7413.236 &      ogarde & 400/11000&  -15.10 &   -11.1 & 213 & 0.970 \\ 
20160205 &  7424.237 &      ogarde & 400/11000&  -14.55 &   -13.6 & 209 & 0.978 \\ 
20160710 &  7579.583 &      othizy & 280/11000&  -15.66 &   -11.1 & 211 & 1.015 \\ 
20160717 &  7586.584 &     afavaro & 200/17000&  -15.73 &   -13.3 & 210 & 0.968 \\ 
20160725 &  7594.526 &     afavaro & 200/17000&  -15.25 &   -14.1 & 209 & 1.041 \\ 
20160728 &  7597.533 &      othizy & 280/11000&  -16.05 &   -17.4 & 211 & 1.019 \\ 
20160812 &  7613.455 &      jterry & 280/17000&  -14.73 &   -13.7 & 206 & 0.941 \\ 
20160817 &  7618.464 &      ogarde & 400/11000&  -16.28 &    -9.5 & 210 & 0.982 \\ 
20160819 &  7619.686 &     tlester & 310/8000 &  -16.50 &    -8.1 & 213 & 0.973 \\ 
20160823 &  7624.487 &    tlemoult & 356/11000&  -16.18 &    -8.9 & 210 & 0.988 \\ 
20160824 &  7625.431 &    adebruin & 280/5800 &  -17.44 &    -9.7 & 215 & 1.005 \\ 
20160825 &  7626.463 &    adebruin & 280/5800 &  -17.36 &    -5.9 & 215 & 1.004 \\ 
20160826 &  7626.548 &      jterry & 280/17000&  -16.67 &    -7.3 & 212 & 1.006 \\ 
20160829 &  7630.426 &    adebruin & 280/5800 &  -17.79 &    -6.5 & 215 & 1.011 \\ 
\hline
\end{tabular}
\end{table*}
\setcounter{table}{0}
\begin{table*}
\caption{Continued.}
\begin{tabular}{l c c c c c c c}
  \hline
  Date & $HJD$ & Observer & tel size/$R$ & $EW$ & $RV$  & width & $V/R$ \\
  (YYYYMMDD) & $-2\,450\,000$ & & (mm)/ & (\AA) & (km\,s$^{-1}$) & (km\,s$^{-1}$) &\\
\hline
20160830 &  7631.431 &    fhoupert & 280/15000&  -18.07 &    -3.3 & 213 & 1.015 \\ 
20160905 &  7637.393 &    adebruin & 280/5800 &  -17.68 &    -8.7 & 215 & 1.010 \\ 
20160906 &  7638.397 &    adebruin & 280/5800 &  -17.21 &    -4.6 & 215 & 1.032 \\ 
20160906 &  7638.401 &      jterry & 280/17000&  -16.73 &    -6.7 & 211 & 1.066 \\ 
20160907 &  7639.402 &    adebruin & 280/5800 &  -17.57 &   -10.3 & 215 & 1.018 \\ 
20160909 &  7641.476 &    fhoupert & 280/15000&  -18.06 &    -1.9 & 213 & 1.029 \\ 
20160912 &  7644.356 &    adebruin & 280/5800 &  -17.57 &    -0.2 & 216 & 1.012 \\ 
20160913 &  7645.387 &    adebruin & 280/5800 &  -17.74 &    -2.0 & 217 & 1.040 \\ 
20160914 &  7646.402 &    adebruin & 280/5800 &  -17.62 &    -4.5 & 215 & 1.034 \\ 
20160924 &  7656.404 &     afavaro & 200/17000&  -17.94 &    -7.6 & 212 & 0.980 \\ 
20160924 &  7656.464 &   epollmann & 400/20000&  -17.40 &    -8.2 & 213 & 1.008 \\ 
20160925 &  7657.356 &   ebertrand & 203/18000&  -18.00 &    -5.9 & 212 & 0.956 \\ 
20160928 &  7660.398 &    fhoupert & 280/15000&  -18.86 &    -8.4 & 214 & 0.976 \\ 
20161002 &  7664.344 &    adebruin & 280/5800 &  -17.90 &    -4.0 & 212 & 0.987 \\ 
20161004 &  7666.388 &      ogarde & 400/11000&  -17.87 &   -11.6 & 210 & 0.966 \\ 
20161004 &  7666.496 &   epollmann & 400/20000&  -17.92 &   -11.6 & 213 & 0.974 \\ 
20161006 &  7668.313 &    adebruin & 280/5800 &  -18.04 &    -7.8 & 215 & 0.964 \\ 
20161010 &  7671.564 &     tlester & 310/8000 &  -17.97 &   -11.9 & 212 & 0.960 \\ 
20161017 &  7679.289 &    adebruin & 280/5800 &  -18.39 &   -15.7 & 216 & 0.988 \\ 
20161020 &  7681.667 &    csawicki & 510/17000&  -17.12 &   -15.2 & 207 & 0.994 \\ 
20161028 &  7690.353 &   ebertrand & 203/13000&  -18.66 &   -13.1 & 213 & 0.970 \\ 
20161030 &  7691.636 &    csawicki & 510/17000&  -17.42 &   -14.5 & 208 & 0.965 \\ 
20161030 &  7692.335 &    fhoupert & 280/15000&  -19.17 &   -11.9 & 213 & 0.958 \\ 
20161031 &  7693.224 &      ogarde & 400/11000&  -18.16 &   -12.4 & 211 & 0.939 \\ 
20161031 &  7693.269 &    adebruin & 280/5800 &  -18.51 &   -10.8 & 216 & 0.975 \\ 
20161107 &  7699.661 &     kgraham & 305/13500&  -18.02 &    -9.6 & 209 & 0.943 \\ 
20161108 &  7700.510 &     tlester & 310/8000 &  -18.19 &   -10.1 & 212 & 0.964 \\ 
20161115 &  7708.362 &      ogarde & 400/11000&  -18.42 &    -5.5 & 211 & 0.966 \\ 
20161124 &  7716.576 &    csawicki & 510/17000&  -17.01 &    -2.9 & 206 & 1.020 \\ 
20161124 &  7717.210 &    adebruin & 280/5800 &  -19.24 &    -6.9 & 216 & 1.008 \\ 
20161126 &  7719.195 &      ogarde & 400/11000&  -18.59 &    -3.1 & 211 & 1.035 \\ 
20161127 &  7720.268 &   epollmann & 400/20000&  -16.14 &   -10.8 & 204 & 1.045 \\ 
20161128 &  7721.206 &    adebruin & 280/5800 &  -19.28 &     1.2 & 217 & 1.023 \\ 
20161129 &  7722.204 &    adebruin & 280/5800 &  -19.23 &     1.3 & 216 & 1.023 \\ 
20161204 &  7727.238 &      ogarde & 400/11000&  -18.60 &    -3.3 & 212 & 1.035 \\ 
20161213 &  7736.266 &      ogarde & 400/11000&  -17.94 &    -5.1 & 211 & 0.993 \\ 
20170726 &  7963.526 &      ctq+mm & 127/12000&  -20.73 &   -11.8 & 212 & 1.015 \\ 
20170731 &  7965.538 &    vdesnoux & 200/15000&  -20.52 &    -9.8 & 211 & 1.045 \\ 
20170805 &  7971.392 &      ctq+mm & 127/12000&  -19.93 &    -6.6 & 208 & 1.038 \\ 
20170806 &  7972.468 &      ogarde & 400/11000&  -20.69 &     0.7 & 209 & 1.071 \\ 
20170814 &  7980.379 &      ctq+mm & 127/12000&  -20.19 &    -5.5 & 211 & 1.022 \\ 
20170817 &  7983.484 &      othizy & 280/11000&  -20.17 &    -4.4 & 209 & 1.053 \\ 
20170826 &  7992.360 &      othizy & 280/11000&  -20.38 &    -6.5 & 210 & 0.984 \\ 
20170828 &  7994.474 &    adebruin & 200/5800 &  -24.45 &    -1.1 & 210 & 0.985 \\ 
20170908 &  8005.378 &      ogarde & 400/11000&  -20.00 &   -15.9 & 208 & 0.975 \\ 
20170909 &  8006.350 &      ctq+mm & 127/12000&  -19.97 &   -11.8 & 211 & 0.996 \\ 
20170919 &  8016.373 &    adebruin & 200/5800 &  -22.39 &   -25.1 & 218 & 1.013 \\ 
20170921 &  8017.708 &     tlester & 310/13000&  -19.51 &   -15.6 & 208 & 1.002 \\ 
20170921 &  8018.371 &    adebruin & 200/5800 &  -20.31 &   -19.8 & 211 & 0.987 \\ 
20170921 &  8018.379 &      othizy & 280/11000&  -20.23 &   -16.8 & 209 & 1.026 \\ 
20170928 &  8025.326 &      othizy & 280/11000&  -20.08 &   -14.8 & 209 & 1.006 \\ 
20170929 &  8026.324 &      ogarde & 400/11000&  -19.79 &    -8.9 & 208 & 0.996 \\ 
20171001 &  8028.290 &      ctq+mm & 127/12000&  -19.85 &   -15.7 & 210 & 1.012 \\ 
20171006 &  8033.280 &      ctq+mm & 127/12000&  -19.97 &   -13.6 & 210 & 0.986 \\ 
\hline
\end{tabular}
\end{table*}
\setcounter{table}{0}
\begin{table*}
\caption{Continued.}
\begin{tabular}{l c c c c c c c}
  \hline
  Date & $HJD$ & Observer & tel size/$R$ & $EW$ & $RV$  & width & $V/R$ \\
  (YYYYMMDD) & $-2\,450\,000$ & & (mm)/ & (\AA) & (km\,s$^{-1}$) & (km\,s$^{-1}$) &\\
\hline
20171006 &  8033.298 &      othizy & 280/11000&  -20.37 &   -13.1 & 209 & 1.011 \\ 
20171009 &  8036.394 &  jguarroflo & 406/9000 &  -19.94 &   -11.1 & 210 & 0.960 \\ 
20171010 &  8037.264 &      othizy & 280/11000&  -20.25 &   -10.0 & 209 & 0.956 \\ 
20171010 &  8037.372 &  jguarroflo & 406/9000 &  -19.95 &   -11.1 & 209 & 0.972 \\ 
20171011 &  8038.375 &  jguarroflo & 406/9000 &  -20.26 &   -11.9 & 210 & 0.966 \\ 
20171011 &  8038.396 &    fhoupert & 280/15000&  -22.04 &    -8.3 & 212 & 0.968 \\ 
20171012 &  8039.373 &  jguarroflo & 406/9000 &  -20.32 &   -10.8 & 210 & 0.976 \\ 
20171012 &  8039.374 &      jterry & 280/17000&  -21.55 &    -1.9 & 211 & 0.956 \\ 
20171013 &  8040.318 &      ogarde & 400/11000&  -20.57 &    -6.7 & 209 & 0.971 \\ 
20171014 &  8041.392 &  jguarroflo & 406/9000 &  -20.12 &    -8.7 & 210 & 0.990 \\ 
20171017 &  8044.357 &  jguarroflo & 406/9000 &  -19.73 &    -7.7 & 209 & 0.989 \\ 
20171020 &  8047.253 &      othizy & 280/11000&  -20.15 &    -6.8 & 208 & 1.032 \\ 
20171020 &  8047.400 &  jguarroflo & 406/9000 &  -20.03 &    -6.7 & 208 & 0.993 \\ 
20171020 &  8047.431 &   ebertrand & 203/13000&  -20.66 &    -1.8 & 210 & 1.007 \\ 
20171022 &  8049.394 &  jguarroflo & 406/9000 &  -19.90 &    -6.5 & 209 & 0.989 \\ 
20171025 &  8052.324 &  jguarroflo & 406/9000 &  -20.19 &    -4.1 & 208 & 0.980 \\ 
20171026 &  8052.618 &     kgraham & 305/13500&  -21.46 &    -3.4 & 211 & 1.027 \\ 
20171027 &  8054.402 &  jguarroflo & 406/9000 &  -20.09 &    -4.4 & 208 & 0.982 \\ 
20171029 &  8056.243 &      ctq+mm & 127/12000&  -20.54 &    -8.1 & 209 & 1.020 \\ 
20171029 &  8056.338 &  jguarroflo & 406/9000 &  -20.21 &    -4.2 & 208 & 1.002 \\ 
20171030 &  8057.321 &    fhoupert & 280/15000&  -21.96 &    -3.7 & 210 & 1.052 \\ 
20171031 &  8058.305 &      ogarde & 400/11000&  -21.42 &    -3.7 & 210 & 1.056 \\ 
20171031 &  8058.370 &  jguarroflo & 406/9000 &  -20.49 &    -4.1 & 209 & 1.060 \\ 
20171101 &  8059.252 &      othizy & 280/11000&  -20.84 &    -5.8 & 208 & 1.076 \\ 
20171101 &  8059.391 &  jguarroflo & 406/9000 &  -20.07 &    -4.1 & 208 & 1.028 \\ 
20171103 &  8061.347 &  jguarroflo & 406/9000 &  -20.34 &    -3.2 & 208 & 1.035 \\ 
20171109 &  8067.326 &  jguarroflo & 406/9000 &  -20.43 &   -12.0 & 209 & 1.042 \\ 
20171111 &  8069.250 &      ctq+mm & 127/12000&  -20.81 &    -3.9 & 210 & 1.040 \\ 
20171113 &  8071.399 &  jguarroflo & 406/9000 &  -19.86 &   -10.7 & 207 & 1.023 \\ 
20171114 &  8072.277 &      othizy & 280/11000&  -21.14 &    -7.9 & 211 & 1.042 \\ 
20171115 &  8073.245 &      othizy & 280/11000&  -20.95 &    -8.0 & 210 & 1.048 \\ 
20171116 &  8074.358 &  jguarroflo & 406/9000 &  -20.31 &   -10.7 & 209 & 1.016 \\ 
20171121 &  8079.347 &  jguarroflo & 406/9000 &  -19.95 &   -14.0 & 207 & 0.978 \\ 
20171122 &  8080.212 &      ogarde & 400/11000&  -21.55 &    -9.6 & 212 & 0.991 \\ 
20171123 &  8081.214 &      othizy & 280/11000&  -20.57 &    -9.9 & 209 & 0.998 \\ 
20171124 &  8081.575 &    csawicki & 510/17000&  -18.97 &   -10.1 & 204 & 0.991 \\ 
20171125 &  8083.373 &  jguarroflo & 406/9000 &  -20.28 &   -11.5 & 207 & 0.989 \\ 
20171128 &  8086.365 &  jguarroflo & 406/9000 &  -20.02 &   -11.5 & 208 & 0.993 \\ 
20171130 &  8088.380 &  jguarroflo & 406/9000 &  -20.41 &   -12.0 & 208 & 1.005 \\ 
20171204 &  8092.209 &      othizy & 280/11000&  -21.01 &   -17.0 & 209 & 1.004 \\ 
20171204 &  8092.350 &  jguarroflo & 406/9000 &  -19.88 &   -18.6 & 207 & 1.016 \\ 
20171212 &  8100.214 &      ogarde & 400/11000&  -21.40 &   -13.7 & 211 & 1.032 \\ 
20171216 &  8104.224 &      ctq+mm & 127/12000&  -20.64 &   -21.1 & 210 & 1.018 \\ 
20171218 &  8106.351 &  jguarroflo & 406/9000 &  -20.12 &   -14.2 & 208 & 1.011 \\ 
20180108 &  8126.580 &    csawicki & 510/17000&  -20.82 &    -8.3 & 208 & 0.964 \\ 
20180114 &  8133.225 &    fhoupert & 280/15000&  -22.46 &    -3.9 & 211 & 1.009 \\ 
20180616 &  8285.587 &      othizy & 280/11000&  -21.61 &   -10.8 & 212 & 0.973 \\ 
20180716 &  8315.539 &      othizy & 280/11000&  -21.84 &    -7.2 & 213 & 1.172 \\ 
20180721 &  8321.459 &  ebryssinck & 280/9500 &  -22.04 &    -0.3 & 213 & 1.166 \\ 
20180727 &  8327.444 &      ctq+mm & 127/12000&  -22.73 &    -6.7 & 216 & 1.105 \\ 
20180810 &  8341.485 &      ogarde & 400/11000&  -21.85 &   -11.8 & 209 & 1.061 \\ 
20180812 &  8342.532 &    fhoupert & 280/15000&  -23.72 &   -15.3 & 212 & 1.053 \\ 
20180818 &  8348.539 &    fhoupert & 280/15000&  -24.06 &   -16.9 & 211 & 1.075 \\ 
20180822 &  8352.516 &      trodda & 235/600  &  -21.47 &     9.6 & 245 & -     \\ 
\hline
\end{tabular}
\end{table*}
\setcounter{table}{0}
\begin{table*}
\caption{Continued.}
\begin{tabular}{l c c c c c c c}
  \hline
  Date & $HJD$ & Observer & tel size/$R$ & $EW$ & $RV$  & width & $V/R$ \\
  (YYYYMMDD) & $-2\,450\,000$ & & (mm)/ & (\AA) & (km\,s$^{-1}$) & (km\,s$^{-1}$) &\\
\hline
20180822 &  8353.393 &      ctq+mm & 127/12000&  -24.04 &   -12.5 & 214 & 1.058 \\ 
20180825 &  8356.512 &    fhoupert & 280/15000&  -24.32 &   -16.0 & 211 & 1.094 \\ 
20180826 &  8357.395 &      ogarde & 400/11000&  -21.92 &   -17.1 & 206 & 1.152 \\ 
20180908 &  8369.533 &    fhoupert & 280/15000&  -25.28 &   -11.8 & 211 & 1.071 \\ 
20180910 &  8372.393 &    adebruin & 200/5800 &  -23.45 &   -10.0 & 210 & 1.023 \\ 
20180910 &  8372.468 &     afavaro & 200/17000&  -16.93 &    -6.4 & 205 & 1.036 \\ 
20180913 &  8375.304 &    adebruin & 200/5800 &  -23.32 &   -15.7 & 209 & 1.034 \\ 
20180915 &  8377.391 &    fhoupert & 280/15000&  -24.98 &    -8.3 & 209 & 1.047 \\ 
20180917 &  8379.353 &      ctq+mm & 127/12000&  -24.55 &    -1.3 & 212 & 1.073 \\ 
20180917 &  8379.407 &    fhoupert & 280/15000&  -24.81 &    -7.6 & 208 & 1.049 \\ 
20180917 &  8379.427 &      othizy & 280/11000&  -23.50 &    -8.9 & 207 & 1.058 \\ 
20180919 &  8381.416 &      othizy & 280/11000&  -23.45 &    -9.8 & 207 & 1.061 \\ 
20180920 &  8382.365 &      othizy & 280/11000&  -23.60 &    -8.8 & 207 & 1.079 \\ 
20180920 &  8382.441 &  jguarroflo & 406/9000 &  -23.23 &    -5.5 & 207 & 1.055 \\ 
20180928 &  8390.474 &  jguarroflo & 406/9000 &  -23.40 &    -6.7 & 206 & 1.065 \\ 
20180929 &  8391.395 &    fhoupert & 280/15000&  -25.28 &    -4.7 & 209 & 1.053 \\ 
20180929 &  8391.414 &      ogarde & 400/11000&  -25.33 &    -3.1 & 212 & 1.048 \\ 
20181003 &  8395.333 &      othizy & 280/11000&  -24.12 &    -6.5 & 207 & 1.091 \\ 
20181003 &  8395.379 &  jguarroflo & 406/9000 &  -22.79 &    -4.5 & 204 & 1.061 \\ 
20181004 &  8396.354 &      othizy & 280/11000&  -23.97 &    -6.4 & 206 & 1.063 \\ 
20181008 &  8400.338 &    adebruin & 200/5100 &  -24.16 &    -7.2 & 207 & 1.078 \\ 
20181008 &  8400.365 &    fhoupert & 280/15000&  -25.81 &    -4.1 & 207 & 1.072 \\ 
20181009 &  8401.273 &    adebruin & 200/5100 &  -24.24 &    -5.9 & 208 & 1.053 \\ 
20181009 &  8401.327 &      othizy & 280/11000&  -24.17 &    -6.5 & 205 & 1.064 \\ 
20181010 &  8402.307 &    adebruin & 200/5100 &  -24.11 &    -2.7 & 208 & 1.108 \\ 
20181011 &  8403.255 &    adebruin & 200/5100 &  -23.82 &    -7.1 & 207 & 1.076 \\ 
20181018 &  8410.304 &    adebruin & 200/17000&  -23.83 &   -16.1 & 208 & 1.031 \\ 
20181018 &  8410.369 &    fhoupert & 280/15000&  -25.15 &    -5.0 & 207 & 1.063 \\ 
20181023 &  8415.322 &    fhoupert & 280/15000&  -25.31 &    -6.8 & 209 & 1.007 \\ 
20181023 &  8415.407 &  jguarroflo & 406/9000 &  -23.51 &    -8.7 & 207 & 1.025 \\ 
20181024 &  8416.298 &      othizy & 280/11000&  -23.88 &    -9.1 & 207 & 1.009 \\ 
20181025 &  8416.540 &     kgraham & 305/12000&  -20.82 &    -7.5 & 198 & 1.017 \\ 
20181102 &  8425.413 &  jguarroflo & 406/9000 &  -23.31 &   -12.3 & 205 & 0.963 \\ 
20181103 &  8426.295 &    fhoupert & 280/15000&  -25.54 &   -11.0 & 209 & 0.970 \\ 
20181108 &  8430.961 &    tbohlsen & 280/14000&  -25.04 &    -8.3 & 208 & 0.945 \\ 
20181110 &  8433.332 &  ebryssinck & 280/15000&  -25.14 &   -12.3 & 205 & 0.921 \\ 
20181112 &  8434.986 &    tbohlsen & 280/14000&  -24.39 &   -12.8 & 208 & 0.950 \\ 
20181114 &  8437.242 &    adebruin & 280/17000&  -22.20 &   -14.6 & 204 & 0.923 \\ 
20181114 &  8437.260 &      othizy & 280/11000&  -23.84 &   -15.1 & 206 & 0.962 \\ 
20181115 &  8438.277 &  ebryssinck & 280/15000&  -24.23 &   -13.5 & 207 & 0.934 \\ 
20181116 &  8439.246 &      othizy & 280/11000&  -23.45 &   -15.3 & 205 & 0.944 \\ 
20181116 &  8439.273 &      ctq+mm & 127/12000&  -25.31 &   -10.3 & 211 & 0.929 \\ 
20181116 &  8439.304 &    fhoupert & 280/15000&  -25.26 &   -12.5 & 208 & 0.915 \\ 
20181118 &  8441.222 &      othizy & 280/11000&  -23.75 &   -15.7 & 207 & 0.943 \\ 
20181119 &  8441.602 &     jdaglen & 356/13000&  -24.83 &   -14.7 & 209 & 0.937 \\ 
20181119 &  8441.954 &    tbohlsen & 280/14000&  -24.26 &   -13.8 & 208 & 0.956 \\ 
20181126 &  8448.994 &    tbohlsen & 280/14000&  -24.09 &    -7.6 & 210 & 0.954 \\ 
20181126 &  8449.348 &  jguarroflo & 406/9000 &  -23.16 &   -11.9 & 207 & 0.954 \\ 
20181128 &  8450.699 &   astiewing & 280/17000&  -23.86 &   -10.5 & 209 & 0.952 \\ 
20181209 &  8461.609 &   astiewing & 280/17000&  -24.14 &    -5.0 & 209 & 0.905 \\ 
20181210 &  8463.337 &  jguarroflo & 406/9000 &  -23.22 &    -6.0 & 208 & 0.909 \\ 
20181216 &  8469.244 &  jguarroflo & 406/9000 &  -23.32 &    -3.2 & 208 & 0.927 \\ 
20181226 &  8479.264 &      ogarde & 400/11000&  -24.53 &     0.0 & 211 & 0.936 \\ 
20190103 &  8487.214 &      ogarde & 400/11000&  -24.27 &    -0.9 & 213 & 0.970 \\ 
\hline
\end{tabular}
\end{table*}
\setcounter{table}{0}
\begin{table*}
\caption{Continued.}
\begin{tabular}{l c c c c c c c}
  \hline
  Date & $HJD$ & Observer & tel size/$R$ & $EW$ & $RV$  & width & $V/R$ \\
  (YYYYMMDD) & $-2\,450\,000$ & & (mm)/ & (\AA) & (km\,s$^{-1}$) & (km\,s$^{-1}$) &\\
\hline
20190106 &  8490.274 &  jguarroflo & 406/9000 &  -23.21 &    -2.9 & 211 & 0.985 \\ 
20190115 &  8499.274 &  jguarroflo & 406/9000 &  -22.83 &    -6.2 & 211 & 0.948 \\ 
20190121 &  8505.215 &    fhoupert & 280/15000&  -25.33 &    -7.8 & 215 & 0.906 \\ 
20190121 &  8505.265 &  jguarroflo & 406/9000 &  -22.67 &    -7.5 & 211 & 0.915 \\ 
\hline
\end{tabular}
\end{table*}

\section*{Acknowledgements}
Y.N. and G.R. acknowledge support from the Fonds National de la Recherche Scientifique (Belgium), the Communaut\'e Fran\c caise de Belgique, the European Space Agency (ESA) and the Belgian Federal Science Policy Office (BELSPO) in the framework of the PRODEX Programme (contract XMaS). This work has made use of the BeSS database, operated at LESIA, Observatoire de Meudon, France. ADS and CDS were used for preparing this document.

\end{document}